\documentclass[aps,prl,twocolumn,tightenlines,superscriptaddress]{revtex4-1}
\usepackage{epsfig,rotating}
\usepackage{multirow}

\newcommand{\nuc}[2]{\hbox{$^{#1}$#2}}

\begin{document}
\title{Quadrupole collectivity beyond $N=28$:
  Intermediate-energy Coulomb
  excitation of \nuc{47,48}{Ar}}   

\author{R.\ Winkler}
   \altaffiliation{Present address: Los Alamos National Laboratory,
     Los Alamos, NM 87545, USA} 
   \affiliation{National Superconducting Cyclotron Laboratory,
      Michigan State University, East Lansing, Michigan 48824, USA}
\author{A.\ Gade}
   \affiliation{National Superconducting Cyclotron Laboratory,
      Michigan State University, East Lansing, Michigan 48824, USA}
   \affiliation{Department of Physics and Astronomy,
      Michigan State University, East Lansing, Michigan 48824, USA}
\author{T.\ Baugher}
    \affiliation{National Superconducting Cyclotron Laboratory,
      Michigan State University, East Lansing, Michigan 48824, USA}
    \affiliation{Department of Physics and Astronomy,
      Michigan State University, East Lansing, Michigan 48824, USA}
\author{D.\ Bazin}
    \affiliation{National Superconducting Cyclotron Laboratory,
      Michigan State University, East Lansing, Michigan 48824, USA}
\author{B.\,A.\ Brown}
    \affiliation{National Superconducting Cyclotron Laboratory,
      Michigan State University, East Lansing, Michigan 48824, USA}
    \affiliation{Department of Physics and Astronomy,
      Michigan State University, East Lansing, Michigan 48824, USA}
\author{T.\ Glasmacher}
    \affiliation{National Superconducting Cyclotron Laboratory,
      Michigan State University, East Lansing, Michigan 48824, USA}
    \affiliation{Department of Physics and Astronomy,
      Michigan State University, East Lansing, Michigan 48824, USA}
\author{G.\, F.\ Grinyer}
   \altaffiliation{Present Address: Grand Acc\'el\'erateur National
     d'Ions Lourds (GANIL), CEA/DSM-CNRS/IN2P3, Bvd Henri Becquerel,
     14076 Caen, France}
   \affiliation{National Superconducting Cyclotron Laboratory,
      Michigan State University, East Lansing, Michigan 48824, USA}
\author{R.\ Meharchand}
    \altaffiliation{Present address: Los Alamos National Laboratory,
     Los Alamos, NM 87545, USA} 
    \affiliation{National Superconducting Cyclotron Laboratory,
      Michigan State University, East Lansing, Michigan 48824, USA}
    \affiliation{Department of Physics and Astronomy,
      Michigan State University, East Lansing, Michigan 48824, USA}
\author{S.\ McDaniel}
    \affiliation{National Superconducting Cyclotron Laboratory,
      Michigan State University, East Lansing, Michigan 48824, USA}
    \affiliation{Department of Physics and Astronomy,
      Michigan State University, East Lansing, Michigan 48824, USA}
\author{A.\ Ratkiewicz}
    \affiliation{National Superconducting Cyclotron Laboratory,
      Michigan State University, East Lansing, Michigan 48824, USA}
    \affiliation{Department of Physics and Astronomy,
      Michigan State University, East Lansing, Michigan 48824, USA}
\author{D.\ Weisshaar}
    \affiliation{National Superconducting Cyclotron Laboratory,
      Michigan State University, East Lansing, Michigan 48824, USA}
\date{\today}

\begin{abstract}
We report on the first experimental study of quadrupole collectivity
in the very neutron-rich nuclei \nuc{47,48}{Ar} using
intermediate-energy Coulomb excitation. These nuclei are located along 
the path from doubly-magic Ca to collective S and Si isotopes, a critical
region of shell evolution and structural change. The deduced $B(E2)$
transition strengths are confronted with large-scale shell-model
calculations in the $sdpf$ shell using the state-of-the-art SDPF-U and
EPQQM effective interactions. The comparison between experiment and
theory indicates that a shell-model description of Ar isotopes around $N=28$
remains a challenge.    
\end{abstract}

\maketitle

Developing predictive power for the fundamental properties of atomic
nuclei is driving experimental and theoretical 
research worldwide. Single-particle motion in a defined nuclear
potential is one of the crucial building blocks for a comprehensive
picture of these strongly interacting fermionic quantum many-body
systems. Large stabilizing energy gaps occur between groups of
single-particle states at certain, ``magic'' fillings with protons or
neutrons. The nuclear potential and resulting shell 
structure have been established in the valley of stability, however, dramatic
modifications to the familiar ordering of single-particle orbits in
``exotic'' nuclei with a large imbalance of proton and neutron numbers
have been found: new shell gaps develop and conventional magic numbers
break down. Current efforts in nuclear physics are aimed at unraveling
the driving forces behind those structural modifications, which are
most pronounced in neutron-rich species~\cite{Bro01,Sor08,Gad08}.

The neutron magic number $N=28$ has attracted much attention in recent
years. On the neutron-rich side of the nuclear chart, below doubly
magic $\nuc{48}{Ca}_{28}$, the $N=28$ shell closure was shown to
disappear progressively below $Z=20$ in
$\nuc{44}{S}_{28}$~\cite{Gla97} and $\nuc{42}{Si}_{28}$~\cite{Bas07},
however, with conflicting experimental data on
$\nuc{46}{Ar}_{28}$~\cite{Sch96,Gad03,Men10}. Much of the initial
spectroscopic information referenced above stems from measurements
of the energy of the first $2^+$ state, $E(2^+_1)$, and the absolute $B(E2;
0^+_1 \rightarrow 2^+_1) \equiv B(E2 \uparrow)$ quadrupole excitation transition
strength. These
observables  can signal the breakdown or 
persistence of a magic number, with high $E(2^+_1)$ and low $B(E2
\uparrow)$ values at a shell gap (reduced collectivity) and the
reverse in the middle of a shell (collective character or deformation). 

The role of the Ar isotopes around $N=28$ is of great
interest. They are, with $Z=18$, between doubly-magic \nuc{48}{Ca} and
the already collective S isotopes ($Z=16$) on the path to
\nuc{42}{Si}, which has the lowest-energy $2^+_1$ state along the $N=28$
isotonic chain to date. Early intermediate-energy Coulomb excitation
measurements~\cite{Sch96,Gla97} showed that \nuc{40,42,44}{S} are
collective with high $B(E2 \uparrow)$ values while, in two independent
measurements, the $B(E2 \uparrow)$ value for  
\nuc{46}{Ar} was found low as one would expect for a persisting $N=28$
shell gap~\cite{Sch96,Gad03}. Shell-model calculations were unable to
explain the reduced collectivity in \nuc{46}{Ar}~\cite{Num01,Now09}
and rather predict the breakdown of $N=28$ as a magic number and an
onset of collectivity already
in Ar. A recent, marginal-statistics excited-state lifetime
measurement extracted a much higher $B(E2 \uparrow)$ value in
agreement with the shell-model description~\cite{Men10}. 

Here, we present the first study of
quadrupole collectivity in the $N=30, 29$ Ar isotopes \nuc{48,47}{Ar}
with intermediate-energy Coulomb excitation. Measured
$B(E2; 0^+_1 \rightarrow 2^+_1)$ and $B(E2; 3/2^-_1 \rightarrow J)$
values are compared to state-of-the-art shell model calculations, taking into
account possible effects of large neutron excess.               

Little is known about the $N=30$ isotones in even-$Z$ nuclei below
\nuc{50}{Ca}. 
In 2007, \nuc{44}{Si} was proven to be bound~\cite{Tar07}, and in 2009
excited states in \nuc{46}{S} were first observed using in-beam
$\gamma$-ray spectroscopy following nucleon-exchange
reactions~\cite{Gad09}. Excited states in \nuc{48}{Ar} were 
studied in 2008 with deep-inelastic reactions~\cite{Bha08} and in 2009 with
nucleon-exchange reactions~\cite{Gad09}. Rare-isotope beams in this
region of the nuclear chart are typically 
produced by the fragmentation of a primary \nuc{48}{Ca} beam on a
light target. In this scheme, the production of exotic projectiles
with $N=30,29$ necessarily involves neutron pickup and/or nucleon
exchange processes which proceed with small cross sections compared to
reactions that only involve proton or neutron removals. Here we show
that, in spite of the small production cross sections, \nuc{47,48}{Ar}
projectile beams can be produced from a \nuc{48}{Ca} primary beam at
intensities sufficient for $\gamma$-ray tagged, thick target
intermediate-energy Coulomb excitation measurements, allowing for the
study of quadrupole collectivity in these very neutron-rich systems.     
  
The measurements were performed at the Coupled Cyclotron Facility at
NSCL on the campus of Michigan State University. The secondary
projectile beam containing \nuc{47,48}{Ar} was produced from a
140~MeV/u \nuc{48}{Ca} stable primary beam impinging on a
681~mg/cm$^2$ \nuc{9}{Be} production target and separated using a
195~mg/cm$^2$ Al wedge degrader in the A1900 fragment
separator~\cite{a1900}. The total momentum acceptance of the separator
was restricted to 2\%. The resulting rare-isotope beam was composed of
~6.5\% \nuc{48}{Ar} and ~57\% \nuc{47}{Ar}. Typical rates of
150~\nuc{48}{Ar}/second and 1300~\nuc{47}{Ar}/second were delivered to
the experimental end station. 

At the pivot point of NSCL's S800 spectrograph~\cite{s800}, a gold
target of 518~mg/cm$^2$ 
thickness was surrounded by the Segmented Germanium Array (SeGA)
consisting of 32-fold segmented high-purity germanium detectors for
high-resolution $\gamma$-ray spectroscopy~\cite{sega}. The
segmentation allows for 
event-by-event Doppler reconstruction of $\gamma$ rays emitted
in flight by the scattered projectiles. For this, the $\gamma$-ray
emission angle 
entering the Doppler reconstruction is determined from the location of
the detector segment with the largest energy deposited. Sixteen
detectors were arranged in two rings with central angles of 90$^\circ$
and 37$^\circ$ with respect to the beam axis. The 37$^\circ$ ring was
equipped with seven detectors while nine detectors occupied 90$^\circ$
positions.  The photopeak efficiency of the array was measured with a
\nuc{152}{Eu} standard calibration source and corrected for target
absorption and the Lorentz boost of the $\gamma$-ray angular distribution
emitted by nuclei moving at velocities of more than $v/c=0.4$.

\begin{figure}[h]
\includegraphics[width=\columnwidth]{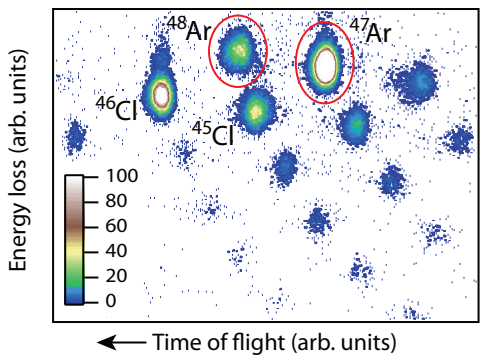}
\caption{\label{fig:pid} (Color online) Particle-identification
  spectrum for the neutron-rich projectile beam passing through the Au
  target. The energy loss measured in the S800 ionization
chamber is plotted versus the ion's flight time. \nuc{47}{Ar} and
\nuc{48}{Ar} can be unambiguously identified among the other species in the
cocktail beam.} 
\end{figure}

Scattered projectiles were identified on an event-by-event basis using the
focal-plane detection system of the large-acceptance S800
spectrograph~\cite{s800}. The ion's energy loss measured in the S800 ionization
chamber and flight-time information measured with two plastic scintillators --
corrected for the angle and momentum of each ion -- were used to unambiguously
identify the scattered projectiles emerging from the target. The identification
spectrum is shown in Fig.~\ref{fig:pid}.

In intermediate-energy Coulomb excitation~\cite{Mot95,Gla98,Gad08},
projectiles are scattered off stable high-$Z$ 
targets and are detected in coincidence 
with the de-excitation $\gamma$ rays, tagging the inelastic
process. Very peripheral collisions are selected in the regime of
intermediate beam energies to exclude nuclear 
contributions to the electromagnetic excitation process. This is
typically accomplished by restricting the data analysis to events at
very forward scattering angles, corresponding to large impact
parameters, $b$, in the interaction between projectile and target
nuclei -- here, we chose $b>1.2A_p^{1/3}+1.2A_t^{1/3}$. Position
measurements from the cathode readout drift chambers in the 
S800 focal plane were combined with ion optics
information to reconstruct the projectile's scattering angle on an
event-by-event basis. The angle-integrated Coulomb excitation
cross section, $\sigma(\theta_{\rm lab} \leq \theta_{\rm lab}^{\rm
  max})\equiv \sigma$, was determined from the
efficiency-corrected $\gamma$-ray intensity relative to the number of
projectiles per number density of the target and translated into
absolute $B(\sigma \lambda)$ excitation strengths using the
Winther-Alder theory of intermediate-energy Coulomb
excitation~\cite{Win79}. 

\begin{figure}[h]
\includegraphics[width=\columnwidth]{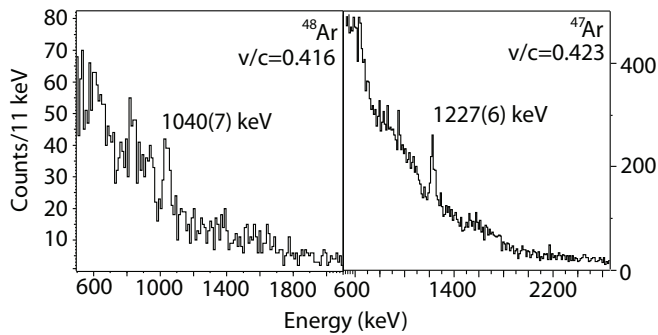}
\caption{\label{fig:gamma} Event-by-even Doppler reconstructed
  $\gamma$-ray spectra detected in coincidence with scattered
  \nuc{48}{Ar} (left) and scattered \nuc{47}{Ar} (right).
Gamma-ray transitions at 1040~keV and 1227~keV can be clearly
identified and are attributed to the de-excitation $\gamma$ rays from
the first excited $2^+$ state in \nuc{48}{Ar} and the first excited
$5/2^-$ state in \nuc{47}{Ar}, respectively.   }

\end{figure}

For the present analysis we use maximum scattering angles of 
$\theta_{\rm lab}^{\rm max}=1.94(5)^{\circ}$ and $1.97(5)^{\circ}$
for \nuc{47}{Ar} (100/u~MeV mid-target energy) and \nuc{48}{Ar} (96/u~MeV mid-target
energy), respectively, corresponding to 
minimum impact parameters just above $b_{min}=13.3$~fm for both
projectile-target systems. The event-by-event
Doppler-reconstructed $\gamma$-ray spectra taken in coincidence
with \nuc{48,47}{Ar} -- with the scattering-angle restrictions applied
-- are shown in Fig.~\ref{fig:gamma}. The $\gamma$-ray transition at
1040~keV  observed in \nuc{48}{Ar} is attributed to the decay of the first
$2^+$ state, in agreement with~\cite{Bha08,Gad09}. The $\gamma$-ray
transition apparent at 1227~keV in the spectrum of \nuc{47}{Ar} is
assigned to the decay of the first excited 
$5/2^-$ state to the $3/2^-$ ground state,
following~\cite{Gau06,Bha08}. The peak-like structure at about 820~keV
in the \nuc{48}{Ar} spectrum is an artefact from Doppler-reconstructed
background lines, only seen in the $37^{\circ}$ ring of SeGA. Similar
structures at the same energy are visible in the spectrum of \nuc{47}{Ar}.     

 
An angle-integrated Coulomb excitation cross section of
$\sigma(\nuc{48}{Ar})=74(11)$~mb was derived for the excitation of the
first $2^+$ state of \nuc{48}{Ar}, translating into $B(E2; 0^+_1 \rightarrow
2^+_1)=346(55)~e^2$fm$^4$ for this most neutron-rich Ar isotope
studied with this technique. The cross section includes
statistical uncertainties and a 5\% uncertainty on the in-beam
$\gamma$-ray detection efficiency. For the extracted $B(E2)$ values, an
additional 5\% systematic 
uncertainty originating from the the
projectile's scattering angle reconstruction was added in quadrature. 

The Coulomb excitation of the odd-$N$ \nuc{47}{Ar} is more
complicated. A mixed 
$E2/M1$ multipolarity is expected for the $5/2^- \rightarrow 3/2^-$
$\gamma$-ray transition. The multipole character of the
radiation affects the $\gamma$-ray angular distribution and thus the
detection efficiency. Large-scale
shell model calculations with different interactions, SDPF-U~\cite{Now09} and
EPQQM~\cite{Kan11}, compute  an identical multipole mixing ratio of
$\delta(E2/M1)=1.01$, corresponding to 50\% $M1$ character. This
$\delta(E2/M1)$ was used in the determination of the in-beam
detection efficiency of SeGA. To put this into perspective,
assuming 100\% $E2$ character would only introduce a relative change of
1.5\% in the detection efficiency. However, in the
excitation process, $E2$ multipolarity  
dominates over $M1$ by orders of magnitude, with 20.9~mb/$100~e^2$fm$^4$
and 0.18~mb/$0.1~\mu_N^2$ for Coulomb excitation via $E2$ and $M1$,
respectively. Shell-model calculations predict $B(M1; 3/2^-
\rightarrow 5/2^-)=0.0236~\mu_N^2$ (SDPF-U)~\cite{Now09} or less
(EPQQM)~\cite{Kan11} and we thus
assume that the excitation proceeds solely via $E2$ character. $B(E2;
3/2^- \rightarrow 5/2^-)=135(17)~e^2$fm$^4$ is deduced from the
measured cross section of $\sigma(\nuc{47}{Ar})=28(3)$~mb, with an error budget as detailed for
\nuc{48}{Ar}. 

\begin{figure}[h]
\includegraphics[width=\columnwidth]{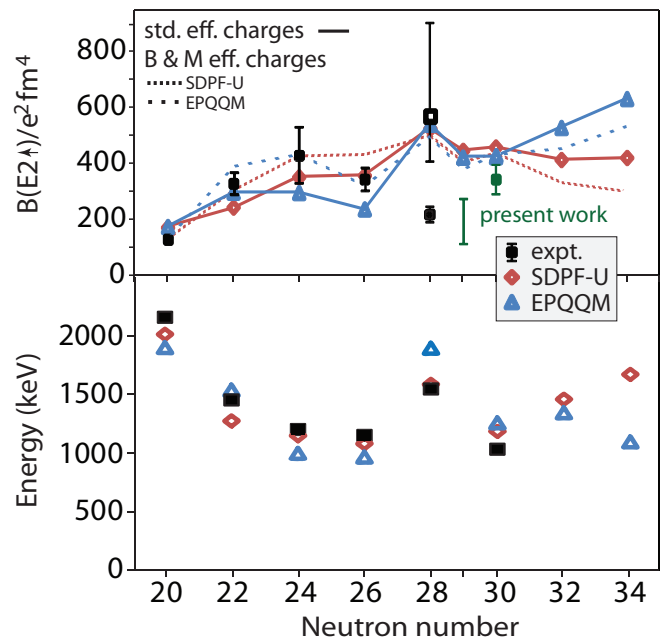}
\caption{\label{fig:evenbe2} (Color online) (upper panel) Measured $B(E2 \uparrow)$ for the
  chain of Ar isotopes~\cite{nndc,Sch96,Gad03,Men10} compared to shell-model
  calculations using two 
  different effective interactions with the standard effective
  charges, $e_p=1.5e$ and $e_n=0.5e$ (solid lines) and $N,Z$-dependent
  effective charges following Bohr and Mottelson (dashed
  lines). (lower panel) Measured and calculated $E(2^+_1)$ energies
  for the Ar isotopic chain.}
\end{figure}

The consistent
description of the onset of collectivity at $N=28$ in the
isotopic chains of sulfur ($Z=16$) and silicon ($Z=14$) has been a
challenge for large-scale shell-model calculations 
and was accomplished by Nowacki and Poves by
devising two effective interactions (SDPF-U) for the $sdpf$ model space, one
valid for $Z \leq 14$ and one to be applied for $Z > 14$~\cite{Now09}. In
a recent shell-model work by Kaneko {\it et al.}, the extended pairing plus
quadrupole-quadrupole force with inclusion of a monopole interaction
(EPQQM) was proposed to provide a consistent description of the
breakdown of $N=28$ across the $Z=20$ to $Z=14$ isotopic
chains~\cite{Kan11}. Both shell-model interactions use
$e_p=1.5e$ and $e_n=0.5e$ for the effective charges that 
enter the computation of $B(E2)$ transition strengths from the proton
and neutron shell-model transition amplitudes $A_p$ and $A_n$ via
$B(E2;J_i \rightarrow J_f)=(e_nA_n+e_pA_p)^2/(2J_i+1)$. In
Fig.~\ref{fig:evenbe2} (upper panel), the shell
model $B(E2;0^+_1 \rightarrow 2^+_1)$ values calculated with the two
different effective
interactions  -- using the standard effective charges quoted above -- are
compared to the  available measured values in the chain of Ar isotopes,
starting from semi-magic
$\nuc{38}{Ar}_{20}$~\cite{nndc,Sch96,Gad03,Men10}. The two calculations 
describe the trend of the data on the even-even 
isotopes well only if one neglects the two consistent, low $B(E2 \uparrow)$
values at $N=28$~\cite{Sch96,Gad03} and assumes that the
marginal-statistics excited-state lifetime measurement for
\nuc{46}{Ar} is correct -- with small differences in the detailed
trends of the two calculations. For \nuc{48}{Ar} the 
calculation is within 2 sigma of the experimental value for the SDPF-U
effective interaction. The EPQQM
effective interaction differs somewhat in the trend approaching $N=28$, with
a marked drop in the $B(E2 \uparrow)$ strength at $N=26$  that is not
present in the SDPF-U results but consistent with the data. For
\nuc{48}{Ar}, the EPQQM calculation 
is within 1.5 sigma of the measured value. The trend beyond $N=30$ is
predicted very differently by the two interactions, with rather
constant $B(E2 \uparrow)$ values out to $N=34$ for the SDPF-U interaction and a
rise in collectivity for EPQQM, with $\nuc{52}{Ar}_{34}$
being the most collective in the chain. The calculated $E(2^+_1)$
energies reproduce the measured energies well where available, with the SDPF-U
and EPQQM predictions diverging at $N=34$. At $N=28$, both
calculations predict high $B(E2 \uparrow)$ and $E(2^+_1)$ values,
deviating from established systematics~\cite{raman} that rather suggests their
anti-correlation as observed for $\nuc{38}{Ar}_{20}$, for example.

The question arises as to the effects of neutron excess on the
transition strength. The effective
charges, which generally compensate 
for missing excitations beyond the (necessarily) limited model space,
can be formulated following Bohr and Mottelson~\cite{BM} to explicitly include
neutron excess via  
$e^{BM}_n=e_{pol}$ and $e^{BM}_p=1+e_{pol}$ where
$e_{pol}=e(Z/A-0.32(N-Z)/A+(0.32-0.3(N-Z)/A)\tau_z)$ is the
polarization charge with $\tau_z=1,-1$ for neutrons and protons,
respectively. This approach approximates the coupling of the particle
motion to higher frequency quadrupole modes that are beyond the model
spaces of the effective shell-model
interactions~\cite{BM}. Figure~\ref{fig:evenbe2} confronts the
experimental data with the shell-model $B(E2 \uparrow)$ values calculated with
$e_{p,n}=1.14~e_{p,n}^{BM}$ (dashed line). A slight renormalization of the
effective charges by 1.14 was introduced to have the results with both
interactions reproduce the $B(E2 \uparrow)$ value of semi-magic 
\nuc{38}{Ar}. With the calculations anchored in this way and the effective
charges evolving with neutron excess, subtle changes in the theory
trends are visible. The description of the available data by the EPQQM
effective interaction improved, in particular below $N=28$, and the
quality of the SDPF-U calculation remains roughly the same. The
differing trend beyond $N=30$ is amplified, with a larger discrepancy
in the prediction of the $B(E2 \uparrow)$ value at $N=34$. If the low
$B(E2 \uparrow)$ value for \nuc{46}{Ar} should be proven correct,
certainly the modified effective charges would not reverse the trend
at $N=28$ and leave a striking disagreement between experiment and
shell model as to the evolution of collectivity below \nuc{48}{Ca}
towards \nuc{42}{Si}.                    
    
With conflicting experimental results at \nuc{46}{Ar} and rather
robust shell-model calculations at $N=28$, it is interesting to turn
to the odd-$N$ 
neighbor \nuc{47}{Ar}. In odd-$A$ nuclei, core-coupled states are
expected to carry the $E2$ strength. These states can be qualitatively described as originating
from a particle or hole weakly coupled 
to the neighboring even-even core~\cite{Ari71}. For \nuc{47}{Ar}, the
coupling of the odd 
$p_{3/2}$ proton to the \nuc{46}{Ar} $2^+_1$ state would give rise to
a quartet of states with spin values of $1/2^-$,
$3/2^-$, $5/2^-$, and $7/2^-$ at about
$E(2^+_1(\nuc{46}{Ar}))=1555$~keV~\cite{Gad03} and with a total $E2$ excitation
strength of $\Sigma_J B(E2; 3/2^-
\rightarrow J)=B(E2\uparrow)_{\nuc{46}{Ar}}$. Indeed, in
both shell model calculations, 
the lowest-lying $1/2^- - 7/2^-$ excited states are found between 1139
- 2125~keV and 950 - 2343~keV in the SDPF-U and EPQQM calculations,
respectively. As one would expect in this extreme weak-coupling
scheme, 85.0\% and 
99.7\% of the shell-model $B(E2 \uparrow)_{\nuc{46}{Ar}}$ are
exhausted by this multiplet in the respective interactions.  

\begin{figure}[h]
\includegraphics[width=\columnwidth]{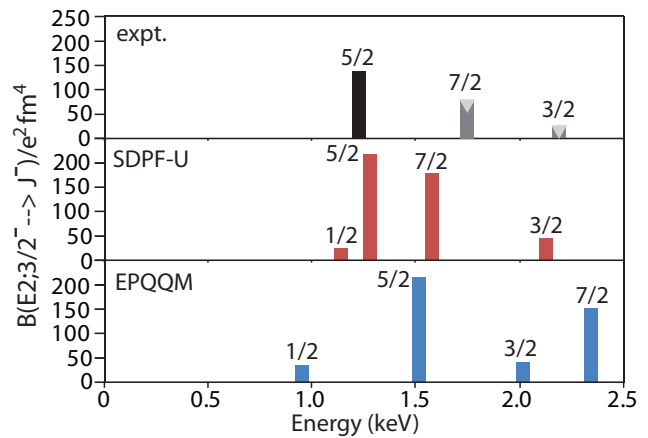}
\caption{\label{fig:odd} $B(E2; 3/2^- \rightarrow J^-)$ strength
  distribution for the lowest-lying excited $1/2^- -
  7/2^-$ states. The experimental values for $J=7/2$ and
  $3/2$ are upper limits as described in the text. Standard
  effective charges were used in both shell-model calculations.}
\end{figure}

In the present measurement, there is only clear evidence
  for one strongly excited state, at 1227~keV with alleged $5/2^-$
  spin assignment~\cite{Gau06,Bha08}. Using the \nuc{47}{Ar} level
  scheme presented in 
  ~\cite{Gau06,Bha08}, and assuming that the observed 1747~keV and 2190~keV
  states are the $7/2^-$ and $3/2^-$ members of the multiplet, we can
  establish upper limits for the $B(E2)$ excitation strengths in the
  present work (we neglect the $1/2^-$ member of the multiplet as this
  state is expected to be only weakly excited). From the
  $\gamma$-ray spectrum of \cite{Bha08}, we roughly estimated 60\% and 40\%
  branches to the ground state and the alleged $5/2^-$
  first excited state, respectively, for both states. We 
  estimated a possible maximum number of $\gamma$-ray counts in the
  expected transitions to the ground state and -- taking into account
  the assumed branching ratios -- extract  
  upper limits of $\sigma(7/2^-) \leq 16(8)$~mb and 
  $\sigma(3/2^-) \leq 4(2)$~mb with conservative 50\%
  uncertainties, translating into generous upper limits of $B(E2;
  3/2^- \rightarrow 7/2^-)<74(37)~e^2$fm$^4$ and   $B(E2; 3/2^- \rightarrow
  3/2^-)<20(10)~e^2$fm$^4$ for the $E2$ excitation strength. In
  Fig.~\ref{fig:odd}, the measured $B(E2)$ values and limits are
  compared to the two shell model calculations (with standard
  effective charges). First, this comparison supports the assignment of
  the level at 1227~keV as the $5/2^-$ state, which -- in both calculations
  -- is predicted to carry the largest fraction of the $E2$ excitation
  strength. We note that the EPQQM interaction predicts the centroid of
  the $E2$ strength to lie about 500~keV higher than in the SDPF-U
  calculation.        

The total $\Sigma_J B(E2; 3/2^- \rightarrow J^-)$ strength
concentrated in these states is included in the $B(E2)$ systematics of
  Fig.~\ref{fig:evenbe2} at $N=29$ for experiment as well as
  theory. The comparison shows that both  
  shell-model effective interactions significantly overpredict the
  $B(E2)$ strength  concentrated in the low-lying states of
  \nuc{47}{Ar}. The experimental $B(E2)$ range given in the figure marks the
  possible span of the total $B(E2)$ strength ranging from the sole
  excitation of 
  the $5/2^-$ states to the  maximum strength observed in case the upper
  limits for the excitation of the $7/2^-$ and $3/2^-$ members of the
  quartet should be realized. This discrepancy is consistent  with the
  situation at $N=28$ if the measured low $B(E2 \uparrow)$  value for
  \nuc{46}{Ar} should be proven correct. This result for \nuc{47}{Ar}
  indicates that challenges remain for the most modern
  shell-model effective
  interactions in this area of rapid structural change and it also 
  demonstrates the seldom exploited, sensitive benchmark posed by
  measures of collectivity in odd-$A$ nuclei.

In summary, we studied the quadrupole collectivity in the very
neutron-rich nuclei \nuc{47,48}{Ar}
via intermediate-energy Coulomb excitation. While the two most recent
effective shell-model effective interactions predict $B(E2 \uparrow)$
values for \nuc{48}{Ar} close to the experimental result, the
low-lying quadrupole collectivity in \nuc{47}{Ar} is significantly
overpredicted by theory, reminiscent of the situation in \nuc{46}{Ar}
if the lower of the conflicting $B(E2 \uparrow)$ values in the
literature should be proven correct. We show that modified effective
charges that approximate effects of neutron excess do not resolve the
discrepancy at $N=28,29$ in the  Ar isotopes and demonstrate
the potential of studies of quadrupole collectivity in odd-$A$ nuclei
to sensitively probe nuclear structure calculations.

\begin{acknowledgments}
This work was supported by
the National Science Foundation under Grants No. PHY-0606007, PHY-1102511 and
PHY-1068217. A.~G. is supported by the Alfred P. Sloan Foundation.
\end{acknowledgments}


\begin{thebibliography}{10}
\bibitem{Bro01} B. A.\ Brown, Prog.\ Part.\ Nucl.\ Phys. 47, 517 (2001).
\bibitem{Sor08} O.\ Sorlin and M.-G.\ Porquet,
  Prog.\ Part.\ Nucl.\ Phys.\ 61, 602 (2008). 
\bibitem{Gad08} A.\ Gade and T.\ Glasmacher, Prog.\ Part.\ Nucl.\ Phys.\ 60,
  161 (2008). 
\bibitem{Gla97} T.\ Glasmacher {\it et al.}, Phys.\ Lett.\ 395, 163 (1997).
\bibitem{Bas07} B.\ Bastin {\it et al.}, Phys.\ Rev.\ Lett.\ 99, 022503
  (2007).
\bibitem{Sch96} H.\ Scheit {\it et al.}, Phys.\ Rev.\ Lett.\ 77, 3967
  (1996).  
\bibitem{Gad03} A.\ Gade {\it et al.}, Phys.\ Rev.\ C 68, 014302 (2003).
\bibitem{Men10} D.\ Mengoni {\it et al.}, Phys.\ Rev.\ C 82, 024308
  (2010). 
\bibitem{Num01} S.\ Nummela {\it et al.}, Phys.\ Rev.\ C 63, 044316 (2001).
\bibitem{Now09} F.\ Nowacki and A.\ Poves, Phys.\ Rev.\ C 79,014310
  (2009). 
\bibitem{Tar07} O.\ B.\ Tarasov {\it et al.}, Phys.\ Rev.\ C 75, 064613 (2007).
\bibitem{Gad09} A.\ Gade {\it et al.}, Phys.\ Rev.\ Lett.\ 102, 182502
  (2009).  
\bibitem{Bha08} S.\ Bhattacharyya {\it et al.}, Phys.\ Rev.\ Lett.\ 101,
  032501 (2008).
\bibitem{a1900} D.\ J.\ Morrissey {\it et al.}, Nucl.\ Instrum.\ Methods
  in Phys.\ Res.\ B 204, 90 (2003).
\bibitem{s800} D.\ Bazin {\it et al.}, Nucl.\ Instrum.\ Methods in Phys.\
  Res.\ B 204, 629 (2003).
\bibitem{sega}  W.\ F.\ Mueller {\it et al.}, Nucl.\ Instrum.\ and
  Methods in Phys.\ Res.\ A 466, 492 (2001).
\bibitem{Mot95} T.\ Motobayashi {\it et al.}, Phys.\ Lett.\ B346, 9 (1995). 
\bibitem{Gla98} T.\ Glasmacher, Annu.\ Rev.\ Nucl.\ Part.\ Sci.\ 48, 1
  (1998). 
\bibitem{Win79} A.\ Winther and K.\ Alder, Nucl.\ Phys.\ A319, 518 
(1979).
\bibitem{Gau06} L.\ Gaudefroy {\it et al.}, Phys.\ Rev.\ Lett.\ 97,
  092501 (2006).
\bibitem{Kan11} K.\ Kaneko, Y.\ Sun, T.\ Mizusaki, and M.\ Hasegawa,
  Phys.\ Rev.\ C 83, 014320 (2011).
\bibitem{raman} S.\ Raman, C. W. Nestor, and P. Tikkanen, At.\ Data
  Nucl.\ Data Tables 78, 1 (2001).  
\bibitem{nndc} S.\ Raman, C.\ W.\ Nestor, and P.\ Tikkanen,
  At.\ Nucl.\ Data Tables 78, 1 (2001). 
\bibitem{BM} A.\ Bohr and B. R.\ Mottelson, Nuclear Structure
  (W. A. Benjamin, Massachusetts, 1975), Vol. 2, p. 515.  
\bibitem{Ari71} A.\ Arima and I.\ Hamamoto,
  Annu.\ Rev.\ Nuc.\ Sci.\ 21, 55 (1971).
\end{thebibliography}
\end{document}